# Giant transverse and longitudinal magneto-thermoelectric effect in polycrystalline nodal-line semimetal $Mg_3Bi_2$

Tao Feng[1]†, Panshuo Wang[2]†, Zhijia Han[1], Liang Zhou[2], Wenqing Zhang[2]*, Qihang Liu[2]*, Weishu Liu[1]*

[1] Department of Materials Science and Engineering, Southern University of Science and Technology, Shenzhen 518055, China

[2] Department of Physics and Shenzhen Institute for Quantum Science & Engineering, Southern University of Science and Technology, Shenzhen 518055, China

Email: liuws@sustech.edu.cn, liuqh@sustech.edu.cn, zhangwq@sustech.edu.cn

**Abstract**

Topological semimetals provide new opportunities for exploring new thermoelectric phenomena, because of their exotic and nontrivial electronic structure topology around the Fermi surface. In this study, we report on the discovery of giant transverse and longitudinal magneto-thermoelectric (MTE) effects in $Mg_3Bi_2$, which is predicted to be a type-II nodal-line semimetal in the absence of spin-orbit coupling (SOC). The maximum transverse power factor is 2182 $\mu Wm^{-1}K^{-2}$ at 13.5 K and 6 Tesla. The longitudinal power factor reaches up to 3043 $\mu Wm^{-1}K^{-2}$ at 15 K and 13 Tesla, which is 20 times higher than in a zero-strength magnetic field and is also comparable to state-of-the-art MTE materials. By compensating Mg loss in the Mg-rich conditions for turning carrier concentration, the sample obtained in this work shows a large linear non-saturating magnetoresistance of 940% under a field of 14 Tesla. This is a two-orders-of-magnitude increase with respect to the normal Mg-deficiency $Mg_3Bi_2$ sample. Using density functional calculations, we attribute the underlying mechanism to the parent nodal-line electronic structure without SOC and the anisotropic Fermi surface shape with SOC, highlighting the essential role of high carrier mobility and open electron orbits in moment space. Our work offers a new avenue toward highly efficient thermoelectric materials through the design of Fermi surfaces with special topological electronic structures in novel quantum materials.

# 1. Introduction

Magneto-thermoelectric (MTE) effects include both transverse and longitudinal transport properties, depending on the directions of the magnetic field ($B$), temperature gradient ($gradT$), and voltage drop ($gradV$). Transverse-MTE is known as the Nernst effect[1,2], where a nonzero $gradV$ signal occurs in the $y$-axis if the $gradT$ is on the $x$-axis and $B$ on the $z$-axis, while the longitudinal-MTE constitutes the magneto-Seebeck effect[3,4], where nonzero $gradV$ occurs on the $x$-axis if $gradT$ is on the $x$-axis and $B$ on the $z$-axis. The magnetic field provides a new degree of freedom for turning carrier transport, broadening the scope of the conventional thermoelectric effect, and promising in waste-heat harvesting and solid-state cooling[5–7]. Over the past years, significant progress has been made in the field of conventional thermoelectric materials through nanoscale techniques, by introducing various atomic defects, nano inclusions, and grain boundaries[8–10]. By contrast, the MTE effect is susceptible to defects because the magnetoelectric coupling is weaker than the thermoelectric effect. As a result, most previous studies have focused on the MTE effect in single crystals. Wang et al. found that the power factor of $Cd_3As_2$ single-crystal greatly increased under a perpendicular magnetic field, yielding power factor increase from 2150 $\mu Wm^{-1}K^{-2}$ to 4700 $\mu Wm^{-1}K^{-2}$ around 350 K[11]. By applying a magnetic field, Chen et al. observed performance enhancement in the transverse power factor manipulated by the bipolar effect, which is the main Seebeck coefficient restriction in the power factor, and a transverse power factor of 400 $\mu Wcm^{-1}K^{-2}$ was achieved at 30 K and 10 Tesla in semimetal $Mg_2Pb$ single-crystal[12].

Fundamentally, MTE materials typically require ultrahigh carrier mobility, electron-hole compensation, and a small Fermi surface[13–15]. Hence, topological semimetals are promising MTE materials because of their special topological surface states and low-energy excitation with linear dispersion[4,16]. For example, Weyl semimetals and Dirac semimetals usually exhibit very high mobility owing to their massless Weyl or Dirac quasiparticles[17–19]. Recently, both the Weyl semimetal NbP and Dirac semimetal $Cd_3As_2$ have been found to have high transverse-MTE properties[11,20]. In addition, particle-hole-like symmetry usually exists in both Weyl and Dirac semimetals, in which the linear bands have led to almost equal concentrations of electron and hole carriers. As a result, this could result in non-saturating magnetoresistance (MR) according to the electron-hole compensation mechanism based on the double carrier model[21], as well as potentially a large longitudinal thermopower (refer to Eq. S(5, 6) in Supplementary Information). In addition, the Fermi surface is tiny when the chemical potential is located around the charge-neutral band-crossing point. Therefore, the unique electronic structure of topological semimetals would result in excellent MTE properties under external knobs, such as a magnetic field.

As a new topological semimetal, $Mg_3Bi_2$-based materials exhibit excellent thermoelectric properties near room temperature[22]. Recent first-principles calculations have shown that a type-II nodal line will appear in the $k_x$-$k_y$ plane of the Brillion zone, while a pair of Dirac points could still exist along the $k_z$ axis[23], even when spin-orbit coupling (SOC) is considered. However, only an MR of 3% was reported at 2 K and 9 Tesla in the $Mg_3Bi_2$ single-crystal bulk, indicating that the magnetic response was much weaker than that in other topological semimetals[24]. Zhou et al. conducted measurements on the longitudinal-MTE properties of $Mg_3Bi_2$ films, which also exhibited a weak MR of 0.6% at 2 K and 2 Tesla[25]. Of note, $Mg_3X_2$ (X = Bi, Sb) samples usually have a high concentration of Mg vacancy because of the low formation energy. Therefore, the previously observed weak MR or MTE responses might not be from the intrinsic behavior of type-II nodal line semimetal, but due to the scattering of Mg vacancies.

Herein, we report on the discovery of significantly boosted transverse and longitudinal MTE effects in $Mg_3Bi_2$ polycrystalline bulk samples synthesized under Mg-rich conditions. We obtain giant transverse thermopower and longitudinal thermopower up to 127 μV/K and 176 μV/K, respectively. In response, the maximum transverse and longitudinal power factors reach 2182 μWm$^{-1}$K$^{-2}$ and 3043 μWm$^{-1}$K$^{-2}$, respectively, for the as-fabricated $Mg_3Bi_2$ polycrystalline sample. The measured electron and hole mobilities are 4843 cm$^2$V$^{-1}$s$^{-1}$ and 1393 cm$^2$V$^{-1}$s$^{-1}$, respectively, an order of magnitude higher than the values previously reported for $Mg_3Bi_2$[24,25]. In addition, polycrystalline $Mg_3Bi_2$ clearly shows a linear unsaturated MR up to 940% at 2 K and 14 Tesla. Our theoretical calculations based on density functional theory (DFT) obtain a strongly anisotropic MR in $Mg_3Bi_2$ due to the anisotropic Fermi surface shape, resulting in giant transverse and longitudinal MTE effects. Our work reveals the essential role of Fermi-surface topology in determining magnetic responses, and provides a general strategy to boost the MTE properties of topological materials through defect engineering.

## 2. Defects engineering toward the intrinsic nodal-line semimetal Mg₃Bi₂

Mg₃Bi₂ has a crystalline structure with a trigonal lattice and space group P$\bar{3}$m1 (Figure S1a)[26]. Bi is located in the interior of a single cell, while Mg usually has two positions: the vertex Mg (Mg-1: 0, 0, 0) and the interior of Mg (Mg-2: 0.3333, 0.6667, 0.6300). It has been reported that Mg at the vertex site can be easily lost during the synthesis process, leading to the formation of Mg vacancy defects[27]. Consequently, polycrystalline Mg₃Bi₂ bulks, synthesized by high-temperature processes, usually tend to present p-types, similar to the situation in Mg₃Sb₂[28]. In addition, the Mg vacancy could also scatter the carriers, resulting in limited mobility in previously reported Mg₃Bi₂ single-crystal bulks[24].

To suppress the Mg vacancies and optimize carrier mobility, a serial of Mg₃Bi₂ samples with a nominal composition of Mg$_{3+x}$Bi₂ ($x$ = 0, 0.1, 0.2, 0.3, 0.4, and 0.5) were successfully synthesized under different Mg compensations with nominal composition using mechanical alloying and spark plasma sintering (Table S1). Among the as-fabricated polycrystalline Mg₃Bi₂ samples [indexed as sample #1 ($x$ = 0), #2 ($x$ = 0.1), #3 ($x$ = 0.2), #4 ($x$ = 0.3), #5 ($x$ = 0.4), and #6 ($x$ = 0.5) with different Mg-rich nominal compositions], sample #5, i.e., Mg$_{3.4}$Bi₂, is determined to be closest to the intrinsic state of the Mg₃Bi₂ nodal line semimetal through hall resistivity and thermal conductivity measurements (refer to part II in the Supplementary Information, Figures S1 to S4). The intrinsic state is one of the critical factors for optimizing carrier mobility and intrinsic Fermi surface shape to enhance the MTE[29]. Through the double carrier model for Hall resistivity $\rho_{xy}$[30]

$$\rho_{xy} = \frac{1}{e} \frac{(n_h \mu_h^2 - n_e \mu_e^2) + \mu_h^2 \mu_e^2 B^2 (n_h - n_e)}{(n_h \mu_h + n_e \mu_e)^2 + \mu_h^2 \mu_e^2 B^2 (n_h - n_e)^2} B, \qquad (1)$$

the carrier mobility of the obtained Mg₃Bi₂ in this work is calculated as $\mu_e$ = 4843 cm²V⁻¹s⁻¹ and $\mu_h$=1393 cm²V⁻¹s⁻¹, which is much higher than previously reported single-crystal bulk (450 cm²V⁻¹s⁻¹) and single film Mg₃Bi₂ materials (120 cm²V⁻¹s⁻¹)[24,25].

## 3. Giant transverse and longitudinal magneto-thermoelectric effect

Figure 1 shows the transverse-MTE and longitudinal-MTE properties of the as-fabricated polycrystalline Mg₃Bi₂ (sample #5), with a nominal composition of Mg$_{3.4}$Bi₂. To study the transverse-MTE properties i.e., Nernst effect of the Mg₃Bi₂ sample, we measured the transverse thermopower $S_{xy}$ (i.e., *gradT* in *x*-axis, *gradV* in *y*-axis, and *B* in *z*-axis) at 1, 3, 6, 9, and 13 Tesla at temperatures range from 2 K to 275 K. $S_{xy}$ increases with the increasing magnetic field and reaches a peak value of 127 μV/K at 13 Tesla and 13.5 K (Figure 1a), which is higher than the previously reported values for polycrystalline NbP[20], and even comparable to $Pb_{0.77}Sn_{0.23}Se$ single-crystal in the low magnetic field range[31]. More detailed comparisons are shown in Figure S5. Furthermore, the higher transverse thermopower also results in a higher Nernst coefficient, according to $N_{xy} = S_{xy}/\mu_0 H$ (Figure S6)[32]. The Nernst coefficient of the as-fabricated polycrystalline $Mg_3Bi_2$ gradually decreases with the magnetic field, consistent with other topological materials[33].

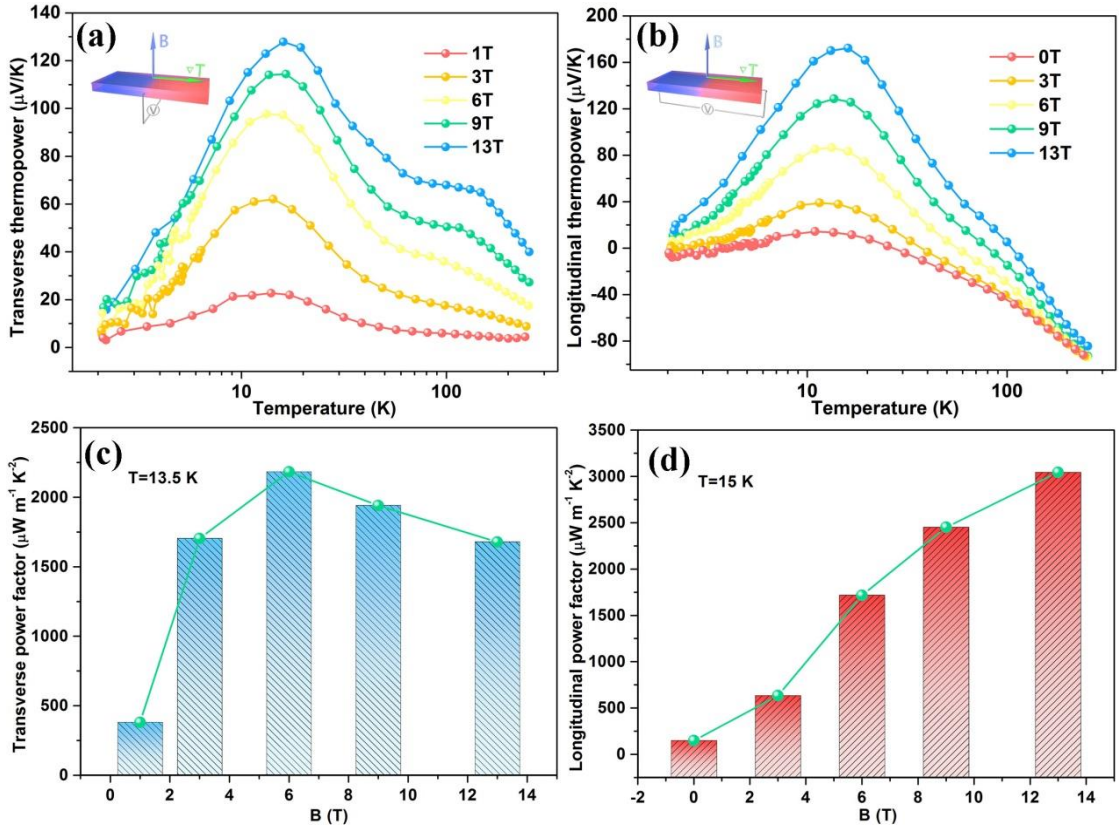

**Figure 1. Transverse- and longitudinal-MTE properties of polycrystalline $Mg_3Bi_2$.** (**A**) The transverse thermopower and (**B**) longitudinal thermopower of $Mg_3Bi_2$ at different magnetic fields in a temperature range of 2 K and 275 K. (**C**) Changes in transverse power factor and (**D**) longitudinal power factor of $Mg_3Bi_2$ with increasing magnetic field at 13.5 K and 15 K, respectively.

For the longitudinal-MTE properties (magneto-Seebeck effect), we measured the temperature-dependent longitudinal thermopower at different magnetic fields. Figure 1b shows the longitudinal thermopower of the $Mg_3Bi_2$ sample at 0, 3, 6, 9, and 13 Tesla in a temperature range of 2 K and 275 K, reaching a maximum of 176 μV/K at about 15 K and 13 Tesla, which corresponded to 13 times enhancement. The longitudinal-MTE effect of $Mg_3Bi_2$ is also sensitive to the Mg-rich conditions (Figure S7a), which is consistent with the carrier mobility and MR. Compared to other MTE materials, the as-fabricated $Mg_3Bi_2$ shows superior gains in longitudinal thermopower under the magnetic field (Figure S7b). Of note, the longitudinal thermopower of the as-fabricated $Mg_3Bi_2$ is still unsaturated up to a magnetic field of 14 Tesla (Figure S8). The observed high and non-saturating longitudinal thermopower in the as-fabricated $Mg_3Bi_2$ should be relative to the special Fermi surface of nodal-line semimetals. Recently, Skinner and Fu theoretically predicted a large and non-saturating thermopower in Dirac/Weyl semimetals subjected to a quantized magnetic field[3], which was supported by the experimental work on Dirac semimetals $Pb_{1-x}Sn_xSe$ and $Cd_3As_2$[11,31]. The longitudinal thermopower feature found in $Mg_3Bi_2$ illustrates that the large and non-saturating thermopower could also be extended to nodal-line semimetals. The transverse power factor $P_{xy}$ and longitudinal power factor $P_{xx}$ can be calculated by[20]

$$P_{xx} = S_{xx}^2 \sigma_{xx}, \qquad (2a)$$

$$P_{xy} = S_{xy}^2 \sigma_{yy}. \qquad (2b)$$

Because the as-fabricated $Mg_3Bi_2$ is a polycrystalline material in this work, the electrical conductivity is isotropic; thus, $\sigma_{xx}$ should be equal to $\sigma_{yy}$. The calculated $P_{xy}$ and $P_{xx}$ for the different magnetic fields are plotted in Figure 1c and 1d, respectively. With increased magnetic field strength, the transverse power factor $P_{xy}$ of $Mg_3Bi_2$ shows a peak at 6 Tesla, while the longitudinal power factor $P_{xx}$ exhibits unsaturated growth. The highest unsaturated $P_{xx}$ is 3043 $\mu Wm^{-1}K^{-2}$ at 15 K under 13 Tesla, which is 20 times higher than the value at zero magnetic field strength. This value is much higher than most previously reported MTE materials, such as polycrystalline NbP (850 $\mu Wm^{-1}K^{-2}$, at 100 K and 9 Tesla) and single crystal $Cd_2As_3$ (2500 $\mu Wm^{-1}K^{-2}$, at 450 K and 7 Tesla)[20,34]. On the other hand, the highest $P_{xy}$ of 2182 $\mu Wm^{-1}K^{-2}$ at 13.5K at 6 Tesla, obtained in the as-fabricated polycrystalline $Mg_3Bi_2$, is also comparable to single-crystal $Pb_{0.77}Sn_{0.23}Se$ (2100 $\mu Wm^{-1}K^{-2}$, at 300 K and 10 Tesla), and close to the value for polycrystalline NdP (3500 $\mu Wm^{-1}K^{-2}$, at 150 K and 9 Tesla)[20,31]. Furthermore, at low-temperature near 15 K, the MTE power factors of $P_{xx}$ and $P_{xy}$ are much higher than conventional thermoelectric materials (Figure S9), suggesting the advantages of MTE materials.

## 4. Giant magnetoresistance effect

The key features of the MTE materials with giant transverse thermopower have been typically attributed to ultrahigh carrier mobility and electron–hole compensation, which are exactly the conditions required for giant magnetoresistance. Owing to the presence of electrons and holes, topological semimetals usually display very high magnetoresistance, which is even not saturating at higher magnetic field strengths of tens of Tesla[35]. Its electrical conductivity $\sigma$ can be significantly modulated under a magnetic field (Part III of Supplementary material), modifying MR and providing a way to improve the longitudinal thermopower (Part I of Supplementary material). Therefore, the magnetoresistance, which reflects the response of electrical transport properties to a magnetic field, is closely related to the MTE effects. Figure 2a compares the MR of the as-fabricated polycrystalline $Mg_3Bi_2$ series materials with nominal composition $Mg_{3+x}Bi_2$ ($x =$ 0, 0.1, 0.2, 0.3, 0.4, and 0.5) at 2 K and 7 Tesla. More detailed MR measurements are shown in Figure S10. All of our samples had positive MR values between 2 and 300 K. Of note, the MR of sample #1 with a nominal composition of $Mg_3Bi_2$ is very close to previously reported $Mg_3Bi_2$ materials in the literature[24,36]. It suggested, in $Mg_3Bi_2$, the coupling between the electric transport and magnetic field is highly sensitive to Mg vacancies. The MR of $Mg_3Bi_2$ at 2 K and 7 Tesla increases from 4%, to 250%, 350%, 410%, 450%, and 413% as the nominal Mg-rich content $x$ increased from $x$=0, to 0.1, 0.2, 0.3, 0.4, and 0.5, respectively. Sample #5, with a nominal composition of $Mg_{3.4}Bi_2$, shows a recording value of 940% at 2 K and 14 Tesla, which is two orders of magnitude higher than previously reported $Mg_3Bi_2$ single crystals[24,25]. It should be noted that the MR is still unsaturated even at a magnetic field strength of up to 14 Tesla. Additionally, excessive Mg results in decreased MR (sample #6, nominal composition of $Mg_{3.5}Bi_2$), further suggesting that sample #5 is closest to the intrinsic state of a semimetal.

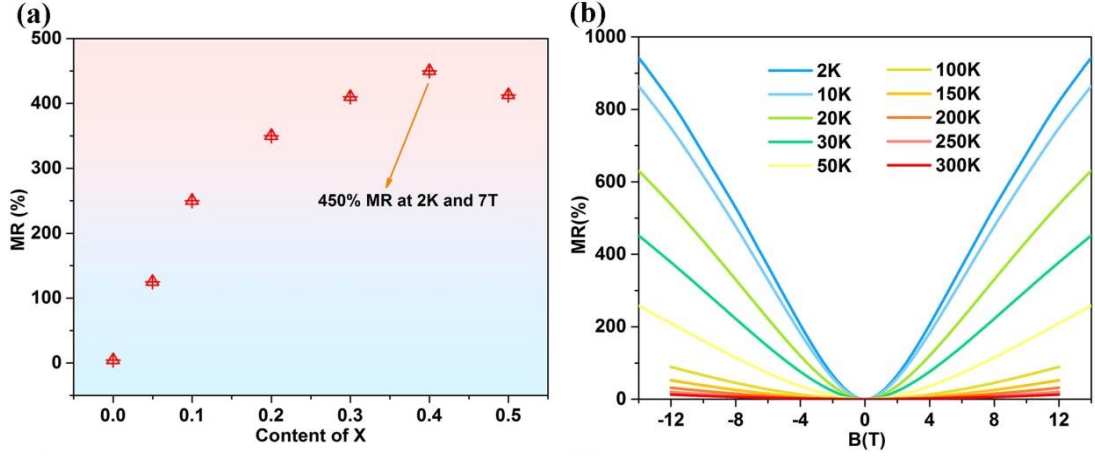

**Figure 2. Magnetoresistance properties of polycrystalline Mg$_3$Bi$_2$.** (**A**) The magnetoresistance at 2 K and 7 Tesla of the as-fabricated polycrystalline Mg$_3$Bi$_2$ with a nominal composition of Mg$_{3+x}$Bi$_2$ ($x$ = 0, 0.1, 0.2, 0.3, 0.4, 0.5) (Table S1). (**B**) The magnetoresistance of sample #5 with a nominal composition of Mg$_{3.4}$Bi$_2$ at different temperatures in a magnetic field up to 14 Tesla.

## 5. Anisotropic magnetoresistance

To understand the anomalous behavior of MR, we performed DFT electronic structure calculations as well as DFT-based transport calculations to simulate the variance of MR as a function of the external magnetic field along the $x$ and $z$ directions. Without considering SOC, the band crossing is close to the Fermi level, showing the type-II nature of the nodal line in Mg$_3$Bi$_2$ (Figure S11). When including SOC, a small energy gap ($\approx$ 40 meV, as shown in Figure 3a) at the nodal points is opened. The corresponding electrical conductivity tensor $\sigma$ of Mg$_3$Bi$_2$ was obtained by applying the following formula[37,38]:

$$\sigma = \sum_n \sigma^{(n)} = \sum_n \frac{e^2}{4\pi^3} \int d\boldsymbol{k}\, \tau[\epsilon_n(\boldsymbol{k})] \boldsymbol{v}_n(\boldsymbol{k}) \overline{\boldsymbol{v}}_n(\boldsymbol{k}) \left(-\frac{\partial f}{\partial \epsilon}\right)_{\epsilon=\epsilon_n(\boldsymbol{k})}, \qquad (3)$$

where $e$ is the electron charge, $\epsilon_n(\boldsymbol{k})$ is the $nth$ band energy at momentum $\boldsymbol{k}$ in the Brillouin zone with the corresponding relaxation time $\tau[\epsilon_n(\boldsymbol{k})]$, $\overline{\boldsymbol{v}}_n(\boldsymbol{k})$ is the weighted average of the Fermi velocity $\boldsymbol{v}_n(\boldsymbol{k})$, and $f$ is the Fermi distribution function of the equilibrium state. The total conductivity $\sigma$ is a summation of the individual band contribution $\sigma^{(n)}$; the MR, i.e., the resistivity tensor $\rho$, can be obtained by inverting the conductivity tensor $\sigma$ (i.e., $\rho = \sigma^{-1}$). Eq. (3) was calculated based on the tight-binding model by Wannier representation extracted from DFT wavefunctions[39,40] (Supplementary Part III). Also, a constant relaxation time approximation was adopted[39–41]. The weighted average group velocity $\overline{\boldsymbol{v}}_n(\boldsymbol{k})$ reflects the nonlinear influence of the magnetic field on electrical conductivity. A larger magnetic field involves more $\boldsymbol{k}$ points' average (Eq. S(8)), leading to saturating or non-saturating MR behavior.

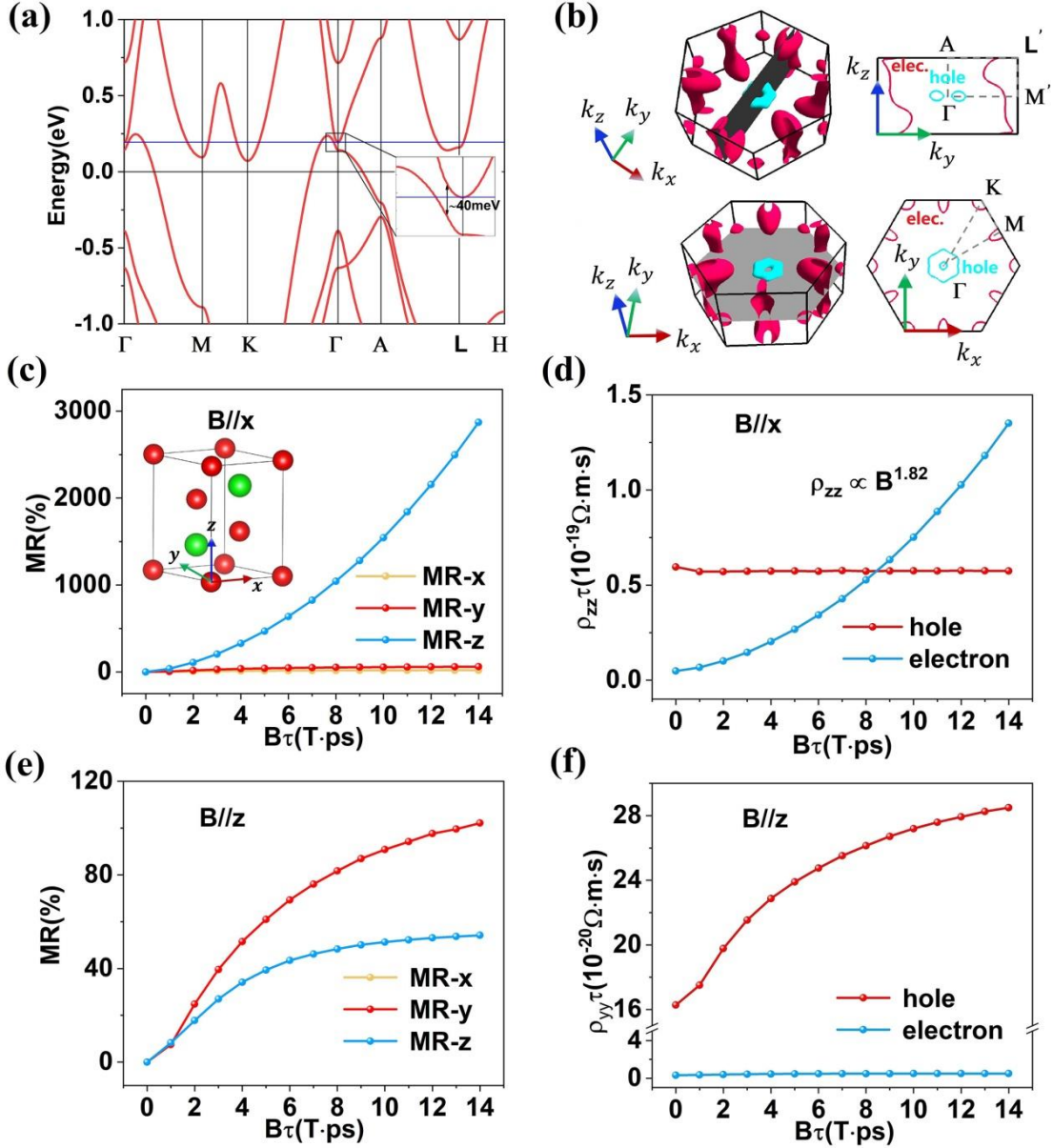

**Figure 3. Electronic structure and transport properties of Mg$_3$Bi$_2$ from first-principles calculations.** (**A**) Band structure of Mg$_3$Bi$_2$ with SOC, where the blue line is the approximated Fermi level for Mg$_3$Bi$_2$ within the rigid band model. (**B**) Fermi surface of Mg$_3$Bi$_2$ (left panel). The cross-section of the Fermi surface was produced by the $k_x = 0$ (upper right) and $k_z = 0$ (bottom right) planes, respectively, exhibiting 'open' orbits of electrons along $k_z$ and 'closed' orbits of holes within the $k_y - k_z$ plane (i.e., $B//x$), and the 'closed' orbits of both electrons and holes in the $k_x - k_y$ plane (i.e., $B//z$). (**C–F**) Field dependence of MR along the $x$ (yellow), $y$ (red), and $z$ (blue) directions with (**C**) $B//x$ and (**E**) $B//z$, respectively. The inset in (**C**) shows the crystal structure. Field dependence of separate contributions of electrons (blue) and holes (red) on the electrical resistivities (**D**) $\rho_{zz}$ for $B//x$ and (**F**) $\rho_{yy}$ for $B//z$.

The calculated transport results are shown in Figure 3c–3f. Here we set the Fermi level $E_f$ = 0.19 eV (i.e., blue line in Figure 3a), corresponding to the as-fabricated sample #5 Mg$_3$Bi$_2$ obtained in this work, where the carrier mobility is optimal because of the change in carrier effective mass (refer to Table S2 in Supplementary Part IV). This is also the band crossing point of the nodal line. First, an anisotropic MR was observed in the hexagonal $Mg_3Bi_2$ when the magnetic field was applied along the *x* axis (Figure 3c). The *z*-axis MR showed a nearly quadratic increase with the magnetic field, while the *y*-axis MR and *x*-axis MR were much smaller and rapidly saturated with ***B*** (Figure 3c). We obtained a giant and unsaturated z-axis MR of near 3000% at $\mathbf{B}\tau = 14$(Tesla·ps) (***B***//*x*). On the other hand, when ***B***//*z* (Figure 3e), MR saturated for all three directions, approaching no more than 105%. Owing to the in-plane rotational symmetry under a perpendicular magnetic field, the MR along the *x* and *y* directions are degenerate.

## 6. Fermi surface topology

To explain the anisotropic MR behavior, we studied the Fermi surface topology of the Bloch electrons and trajectories of the charge carriers driven by the Lorentz force. Figure 3b shows the Fermi-surface contours with $E_f = 0.19$ eV (blue line in Figure 3a). Based on the semiclassical scenario (refer to Eq. (S9) in the Supplementary Part III), the directions of the magnetic field ***B***, the Fermi velocity $v_n(\mathbf{k}(t))$, and the derivative of ***k*** (i.e., $\dot{\mathbf{k}}$) are perpendicular to each other. Thus, the orbits of the Bloch electrons in the Brillouin zone are the cross-sections of the Fermi surface and a plane which is simultaneously normal to ***B*** crossing the $\mathbf{k}(t = 0)$ point. For ***B***//*x*, the Fermi surface for $k_x = 0$ plane is shown in the upper right panel of Figure 3b. The hole orbits near the Γ point are closed circles, while the electron orbits are open, i.e., extending throughout the Brillouin zone along the $k_z$ direction, resulting in fewer electrons moving along the *z* direction compared to the zero magnetic field condition. As a result, with an increase in magnetic field strength along *x*, the average velocity of the electrons along z is significantly suppressed (refer to Eq. (S8) in the Supplementary Part III), resulting in increased resistivity $\rho_{zz}$. To confirm this, we theoretically separated the contributions of the electrons and holes for resistivity $\rho_{zz}$, as shown in Figure 3d. Electron resistivity exponentially increased with ***B**τ* with a power factor of 1.82, while the hole contributions remained nearly unchanged upon magnetic field, consistent with the understandings of orbit topology.

For ***B***//*z*, the orbits of both electrons and holes are closed (bottom right, Figure 3b), leading to the saturation properties of MR for all the directions. The separate contributions of the electrons and holes for $\rho_{yy}$ shown in Figure 3f further confirmed the closed-orbit nature of the Fermi surface. Considering that our experimental sample is polycrystalline, in a first-order approximation that the total MR is an average of all directions, which is consistent with our calculation results.

The above results imply that we could choose the optimal doping level in experiment to obtain large MR as well as MTE by designing the Fermi surface topology. Specifically, the Fermi level at $E_f = 0.19$ eV of $Mg_3Bi_2$ is optimal for nonsaturated MR, where the high carrier mobility and open carrier orbits occur simultaneously. On the other hand, the nonsaturated MR effect will be suppressed as the Fermi level moves away from the optimal value, which is consistent with our MR measurements for various Mg doping levels (Figures 2a and S12). For example, when $E_f = 0.15$ eV, both the electron and hole orbits are almost closed (Figure S13a), leading to a saturated MR for $Mg_3Bi_2$ (Figures S12a and S12b); when Fermi level is further increased to 0.21 eV, more electron orbits around Gamma are closed, not favorable for nonsaturated MR as well (Figures S12c, S12d, and S13c). Of note, MR is small and saturated for the parent electronic structure without SOC at the band crossing point (Figure S14). This indicates that the nodal line semimetal with a tiny energy gap is a promising candidate system for realizing giant MTE.

## 7. Conclusion

In summary, we revealed the giant transverse and longitudinal magneto-thermoelectric effects in a polycrystalline nodal-line semimetal $Mg_3Bi_2$. An extra Mg compensation effectively inhibited the formation of Mg vacancy defects, rendering the sample with a nominal composition of $Mg_{3.4}Bi_2$ very close to the stoichiometric state. Remarkably, polycrystalline $Mg_3Bi_2$ (sample #5, with a nominal composition of $Mg_{3.4}Bi_2$) shows a high transverse thermopower of 127 μV/K and longitudinal thermopower of 176 μV/K at 13 Tesla. As a result, the transverse power factor is 2182 μWm$^{-1}$K$^{-2}$ at 13.5 K, while the longitudinal power factor reaches up to 3043 μWm$^{-1}$K$^{-2}$ at 15 K, 20 times higher than those measured without magnetic field. Meanwhile, the MR of $Mg_3Bi_2$ gradually increases and reached a maximum value of 940% at 14 Tesla and 2 K, indicating the common coherence of MR and MTE effects. First-principles calculations revealed that the unsaturated MR behavior originates from the anisotropic Fermi surface topology of $Mg_3Bi_2$, owing to the hexagonal crystal structure and parent nodal-line electronic structure. With SOC, the nodal line opens a small gap, giving rise to nonsaturating MR as well as a large thermopower. Our findings highlight the great potential of $Mg_3Bi_2$-based materials for energy conversion through the MTE effect, providing a new route for obtaining high-performance MTE materials through the inverse design of Fermi surface topology.

## 8. Experimental Section

*Sample preparation*: The $Mg_{3+x}Bi_2$ samples were synthesized through a combined approach of mechanical alloying and spark plasma sintering (SPS). Magnesium turnings (Mg, 99.98%, Acros Organics) and bismuth shots (Bi, 99.999%, 5N Plus) were weighed according to the composition of $Mg_{3+x}Bi_2$ ($0 \leq x \leq 0.5$), and were loaded into a stainless-steel ball milling jar in a glove box under an Ar atmosphere with an oxygen level of <1 ppm. After ball milling for 10 hours in a SPEX 8000M mixer, the ball-milled powders were loaded into a graphite die with an inner diameter of 15 mm, in the glove box. The graphite die with the loading powder was immediately sintered at 700°C under a pressure of 50 MPa for 5 min via spark plasma sintering (SPS) (SPS-211Lx, Fuji Electronic Industrial Co. LTD). The SPS bulks were ~15 mm in diameter, with a thickness of ~8 mm.

*Sample characterization*: The phase purity of the product was measured by powder X-ray diffraction (XRD) on a Rigaku D/Max-2550 instrument (Cu Kα radiation, $\lambda = 1.5418$ Å, 18 KW). The temperature-dependent thermal conductivity was measured adiabatically using the thermal transport option (TTO) on a Quantum Design Physical Property Measurement System (PPMS 14 Tesla) with one-heater and two-thermometer configuration. With the same option, the transverse and longitudinal thermopower were simultaneously measured adiabatically. We mounted the sample with transverse voltage leads, which were purposefully offset by a distance X from each other so that the longitudinal (Seebeck) and transverse (Nernst) voltage could both be measured. The heat flowed from left to right, and the thermometers T-hot and T-cold were mounted on the left and right leads located on the lower side of the sample. The V+ and V− leads measured the diagonal voltage component, and by measuring it as a function of the magnetic field in both the positive and negative fields, we could employ symmetry arguments to separate the longitudinal thermopower from the transverse thermopower. The longitudinal and Hall resistivities were measured using the electronic transport option (ETO) in PPMS using the standard four-probe method.

*Calculation methods*: Electronic structure calculations were based on the density functional theory (DFT) implemented in the Vienna ab initio simulation package (VASP)[42,43], where the exchange-correlation potential was treated by generalized gradient approximation (GGA) of the Perdew−Burke−Ernzerhof (PBE) functional[44] and the ionic potential was based on the projector augmented wave (PAW) method[45,46]. Owing to the strong relativistic effect in Bismuth, spin-orbit coupling (SOC) was also considered for the energy band dispersion calculations. The energy cutoff of the plane wave was set to 600 eV. For Brillouin zone sampling, a $19 \times 19 \times 9$ Gamma centered K-point mesh was used for the 5-atom unit cell for self-consistent calculations, and an $11 \times 11 \times 5$ mesh was used for structure relaxation. The convergence criteria of the energy and force were set to $10^{-6}\ eV$ and 0.002 eV/Å, respectively. The tight-binding model Hamiltonian adapted for the Wannier interpolation implemented in the WannierTools package[39,40] (i.e., electrical conductivity calculation) was constructed by the Wannier90 software[47] using the maximally localized Wannier function approach[48–50]. The s, p orbits of Mg and p orbits of Bi were selected as the initial projectors for Wannier90.

## Supporting Information

Supporting Information is available from the Wiley Online Library or from the author.

## Acknowledgements

This work was supported by by the National Key R&D Program of China (2019YFA0704900), Shenzhen Science and Technology Basic Research Program (JCYJ20170817111443306) and Shenzhen Science Technology Fund (No. KYDPT20181011104007). The authors would like to thank the support of Core Research Facilities (SCRF) of Southern University of Science and Technology. W.S.L. acknowledges the support from the Tencent Foundation through the XPLORER PRIZE.

**Conflict of Interest:**

The authors declare that they have no competing interests.

**Data Availability Statement**

The data that support the findings of this study are available from the corresponding author upon reasonable request.

**Keywords**

Magneto-Thermoelectric, Magnetoresistance, Fermi surface topology, Polycrystalline $Mg_3Bi_2$

84, 1419.

# Supplementary Material for

# Giant transverse and longitudinal magneto-thermoelectric effect in polycrystalline nodal-line semimetal $Mg_3Bi_2$

Tao Feng[1]†, Panshuo Wang[2]†, Zhijia Han[1], Liang Zhou[2], Wenqing Zhang[2]*, Qihang Liu[2]*, Weishu Liu[1]*

[1]Department of Materials Science and Engineering, Southern University of Science and Technology, Shenzhen 518055, China

[2]Department of Physics and Shenzhen Institute for Quantum Science & Engineering, Southern University of Science and Technology, Shenzhen 518055, China

Email: liuws@sustech.edu.cn, liuqh@sustech.edu.cn, zhangwq@sustech.edu.cn

This PDF file includes:

    Methods (Part I, Part II, and Part III)

    Figs. S1 to S14 (Part IV)

    Table S1 and S2 (Part IV)

    Reference

## Part I. Phenomenological expression of the thermopower S

Generally, the electrical current $\boldsymbol{J}$ can be expressed by [1-3]

$$\boldsymbol{J} = \boldsymbol{\sigma} \cdot \boldsymbol{E} + \boldsymbol{\alpha} \cdot (-\nabla T), \tag{S1}$$

where $\boldsymbol{\sigma}$, $\boldsymbol{E}$, $\boldsymbol{\alpha}$, and $\nabla T$ denote the electrical conductivity tensor, electric field vector, thermoelectric conductivity tensor, and temperature gradient, respectively. For the thermoelectric measurement, an open-circuit condition was used and with no pure current; thus, $\boldsymbol{J}$ is set to 0. Without loss of generality, we considered a case where the temperature gradient is along the $x$ direction (i.e., $\nabla_x T \neq 0$, $\nabla_y T = \nabla_z T = 0$), and the magnetic field $\boldsymbol{B}$ or the magnetic moment $\boldsymbol{M}$ is along the $z$ direction. Then, only the in-plane components (i.e., $E_x$ and $E_y$) of thermoelectric field $\boldsymbol{E}$ is nonzero. Based on these conditions, Eq. (S1) can be simplified as

$$\begin{bmatrix} \sigma_{xx} & \sigma_{xy} \\ \sigma_{yx} & \sigma_{yy} \end{bmatrix} \begin{bmatrix} E_x \\ E_y \end{bmatrix} = \begin{bmatrix} \alpha_{xx} & \alpha_{xy} \\ \alpha_{yx} & \alpha_{yy} \end{bmatrix} \begin{bmatrix} \nabla_x T \\ \nabla_y T \end{bmatrix}. \tag{S2}$$

Combined with the Onsager's relation, that is $\sigma_{xy} = -\sigma_{yx}$ and $\alpha_{xy} = -\alpha_{yx}$, the thermoelectric field can be expressed as

$$\begin{bmatrix} E_x \\ E_y \end{bmatrix} = \begin{bmatrix} \sigma_{xx} & \sigma_{xy} \\ \sigma_{yx} & \sigma_{yy} \end{bmatrix}^{-1} \begin{bmatrix} \alpha_{xx} \\ \alpha_{yx} \end{bmatrix} \nabla_x T = \frac{1}{\sigma_{xx}\sigma_{yy} + \sigma_{xy}^2} \begin{bmatrix} \sigma_{yy}\alpha_{xx} - \sigma_{xy}\alpha_{yx} \\ \sigma_{xy}\alpha_{xx} + \sigma_{xx}\alpha_{yx} \end{bmatrix} \nabla_x T. \tag{S3}$$

According to the definition of thermopower $S$, we can finally get

$$\begin{cases} S_{xx} = \dfrac{E_x}{-\nabla_x T} = -\dfrac{\sigma_{yy}\alpha_{xx} - \sigma_{xy}\alpha_{yx}}{\sigma_{xx}\sigma_{yy} + \sigma_{xy}^2} \\ S_{xy} = \dfrac{E_y}{-\nabla_x T} = -\dfrac{\sigma_{xy}\alpha_{xx} + \sigma_{xx}\alpha_{yx}}{\sigma_{xx}\sigma_{yy} + \sigma_{xy}^2} \end{cases}. \tag{S4}$$

Usually, the transverse electric field $\sigma_{xy}$ will be several orders of magnitude smaller than the longitudinal terms $\sigma_{xx}$ and $\sigma_{yy}$. Thus, the thermopower $S$ can be approximately expressed by

$$\begin{cases} S_{xx} \approx -\dfrac{\sigma_{yy}\alpha_{xx} - \sigma_{xy}\alpha_{yx}}{\sigma_{xx}\sigma_{yy}} \approx -\alpha_{xx}/\sigma_{xx} \\ S_{xy} \approx -\dfrac{\sigma_{xy}\alpha_{xx} + \sigma_{xx}\alpha_{yx}}{\sigma_{xx}\sigma_{yy}} \approx \alpha_{xy}/\sigma_{yy} - S_{xx}(\sigma_{xy}/\sigma_{yy}) \end{cases}. \tag{S5}$$

As shown in Eq. (S5), there are two routes to enhance the thermopower $S$: a larger thermoelectric conductivity tensor $\alpha$ or a smaller longitudinal electric conductivity $\sigma$ (i.e., a larger magnetoresistance). It is worth noting that if $\sigma_{xy}$ could be compared with $\sigma_{xx}$, a more comprehensive analysis would be required and the above simple argument should be reconsidered.

## Part II. Determining the best Mg₃Bi₂ sample

The Mg₃Bi₂ series samples (#1 to 6) were synthesized through combined mechanical alloying and spark plasma sintering (SPS). The nominal compositions of these different Mg₃Bi₂ samples are presented in Table S1. Figure S1b presents the XRD patterns of the obtained Mg₃Bi₂ materials recorded at room temperature. The diffraction peaks of all the samples are well matched with the standard data of trigonal Mg₃Bi₂ (JCPDS No. 65-1909), confirming a pure phase. Next, the thermal conductivities $\kappa_{xx}$ were measured to verify the changes in Mg vacancies with different Mg-rich conditions. Figure S1c shows the temperature-dependent $\kappa_{xx}$ of the as-fabricated Mg₃Bi₂ series of samples, with peaks near 20 K. The lattice thermal conductivity in the low-temperature range is very sensitive to atomic defects; thus, the peak $\kappa_{xx}$ increased from 6.06 Wm$^{-1}$K$^{-1}$ for sample #1 to 19.49 Wm$^{-1}$K$^{-1}$ for sample #5, suggesting that the concentration of Mg vacancies gradually decreases under Mg-rich conditions. However, as the Mg content further increased, i.e., sample #6, the thermal conductivity began to decrease, which should be relevant to the new Mg interstitial defects in the Mg-rich condition [4].

Figure S1d shows the Seebeck coefficient ($S_{xx}$) of the obtained samples at room temperature. The as-fabricated Mg₃Bi₂ sample #1 has a room temperature (RT) $S_{xx}$ of 40 µV/K due to intrinsic Mg vacancies, which is consistent with the previously reported Mg₃Bi₂ (42 µV/K) [5]. Also, $S_{xx}$ near RT for the as-fabricated Mg₃Bi₂ series samples changes from p-type to n-type, indicating that the dominant atomic defect changes from acceptor-like V$_{Mg}$ to donor-like Mg$_i$. We also conducted first-principles calculations to investigate the effect of V$_{Mg}$ (in supercell: Mg$_{2.67}$Bi₂, Mg$_{2.75}$Bi₂) and Mg$_i$ (in the supercell: Mg$_{3.25}$Bi₂) on the band structure (Figure S2). Compared to Mg₃Bi₂, both defects V$_{Mg}$ and Mg$_i$ did not significantly change the band structure; however, we did observe a Fermi energy shift toward the conduction band from Mg$_{2.67}$Bi₂ to Mg$_{2.75}$Bi₂, Mg₃Bi₂, and Mg$_{3.25}$Bi₂, respectively. This result is consistent with the experimental trend of the Seebeck coefficient. Generally, Mg loss due to evaporation is carefully balanced in the as-fabricated polycrystalline Mg₃Bi₂ bulk synthesized under Mg-rich conditions, providing a material basis for investigating the transverse/longitudinal magneto-thermoelectric properties.

Figure S1e and S1f show the electron mobility ($\mu_e$) and hole mobility ($\mu_h$) of the as-fabricated $Mg_3Bi_2$ series polycrystalline bulk, according to the double carrier model of Hall resistivity $\rho_{xy}$ (Figure S3) [6]:

$$\rho_{xy} = \frac{1}{e} \frac{(n_h \mu_h^2 - n_e \mu_e^2) + \mu_h^2 \mu_e^2 B^2 (n_h - n_e)}{(n_h \mu_h + n_e \mu_e)^2 + \mu_h^2 \mu_e^2 B^2 (n_h - n_e)^2} B. \tag{S6}$$

For sample #1, the electron and hole mobilities are 1644 $cm^2V^{-1}s^{-1}$ and 274 $cm^2V^{-1}s^{-1}$, respectively. For the samples (#2 to #6) synthesized under different Mg-rich conditions, the carrier mobility gradually increases and reaches its maximum in samples #5, i.e., $\mu_e$ = 4843 $cm^2V^{-1}s^{-1}$ and $\mu_h$=1393 $cm^2V^{-1}s^{-1}$. Additionally, the ratio of hole to electron concentration at 2 K in the $Mg_3Bi_2$ series samples changes from 227.09, to 6.28, 3.05, 2.61, 2.08, and 2.30, respectively (Figure S4). Balancing the ratio of $n_h/n_e$ is a critical challenge for synthesizing MTE materials. Obviously, the ratio of carrier concentration in sample #5 is closest to 1, indicating that a strong electron-hole compensation effect may exist in this $Mg_3Bi_2$ sample. Finally, after analyzing thermal conductivity, carrier mobility, and the carrier concentration ratio, sample #5 is determined to be the closest to the intrinsic state of the $Mg_3Bi_2$ semimetal.

## Part III. Formula for electrical resistivity with an external magnetic field

With the relaxation time approximation and the Boltzmann equation, we followed Ashcroft and Mermin [7], taking into account the uniform magnetic field, and the electrical conductivity tensor for each band was formulated as

$$\sigma^{(n)} = \frac{e^2}{4\pi^3} \int d\boldsymbol{k}\, \tau[\epsilon_n(\boldsymbol{k})] \boldsymbol{v}_n(\boldsymbol{k}) \bar{\boldsymbol{v}}_n(\boldsymbol{k}) \left(-\frac{\partial f}{\partial \epsilon}\right)_{\epsilon=\epsilon_n(\boldsymbol{k})}, \quad (S7)$$

where e is the element charge, $\epsilon_n(\boldsymbol{k})$ is the $nth$ band energy at the $\boldsymbol{k}$ point in the Brillouin zone with the corresponding relaxation time $\tau[\epsilon_n(\boldsymbol{k})]$, $f$ is the Fermi distribution function of equilibrium state, $\boldsymbol{v}_n(\boldsymbol{k})$ is the Fermi velocity, and $\bar{\boldsymbol{v}}_n(\boldsymbol{k})$ is a weighted average of the velocity over the past history of the electron orbit passing through $\boldsymbol{k}$:

$$\bar{\boldsymbol{v}}_n(\boldsymbol{k}) = \int_{-\infty}^{0} \frac{dt}{\tau_n} e^{\frac{t}{\tau_n}} \boldsymbol{v}_n(\boldsymbol{k}(t)). \quad (S8)$$

The time evolution of the wave vector $\boldsymbol{k}$ followed:

$$\frac{d\boldsymbol{k}_n(t)}{dt} = -\frac{e}{\hbar} \boldsymbol{v}_n(\boldsymbol{k}(t)) \times \boldsymbol{B}. \quad (S9)$$

The total conductivity $\sigma$ consisted of a summation of the individual band contribution $\sigma^{(n)}$:

$$\sigma = \sum_n \sigma^{(n)}. \quad (S10)$$

To further calculate MR, the resistivity tensor ρ could be obtained by inverting the total conductivity tensor σ:

$$\rho = \sigma^{-1}. \quad (S11)$$

**Part IV.**

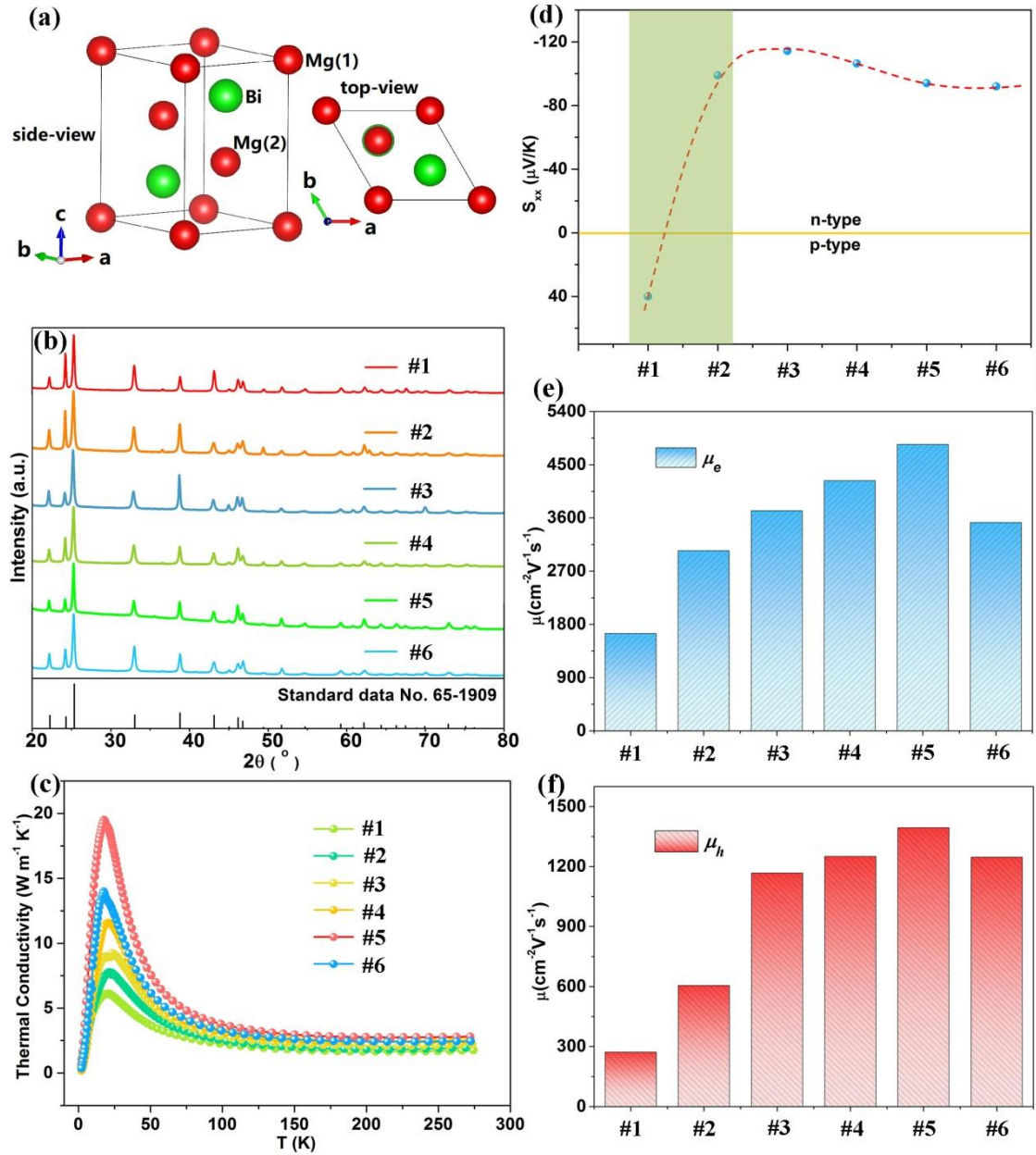

Figure S1. Characterization of the structure, thermal conductivity, Seebeck coefficient, and carrier mobility in the $Mg_3Bi_2$ series samples. (a) The crystal structure of $Mg_3Bi_2$ in the space group $P\bar{3}m1$. (b) XRD patterns. (c) Temperature-dependent thermal conductivity. (d) Seebeck coefficient. (e) Electron mobility, and (f) hole mobility of the obtained $Mg_3Bi_2$ samples #1 to #6.

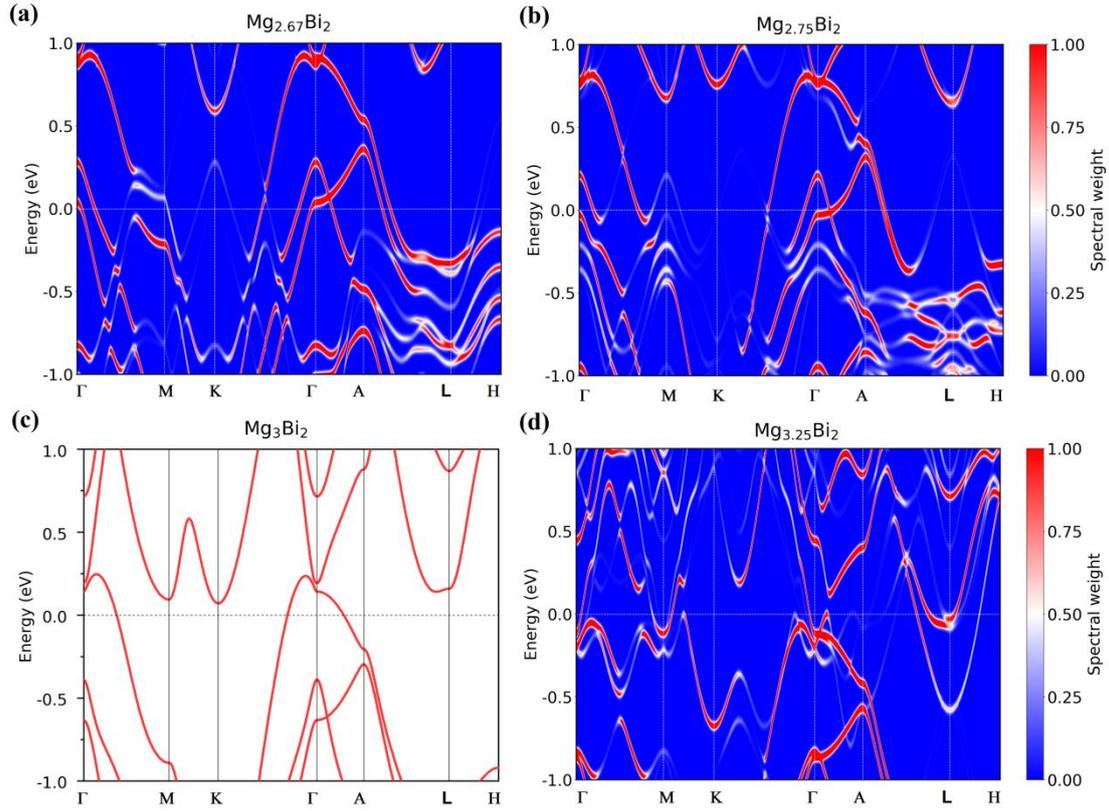

Figure S2. Band structure of Mg$_3$Bi$_2$ (c) and effective band structures of Mg$_{3+x}$Bi$_2$ (a, b, d) with different Mg content with SOC. The main features of the effective band structures around the Fermi level are similar to Mg$_3$Bi$_2$ with effective chemical potential, indicating the applicability of the rigid band model.

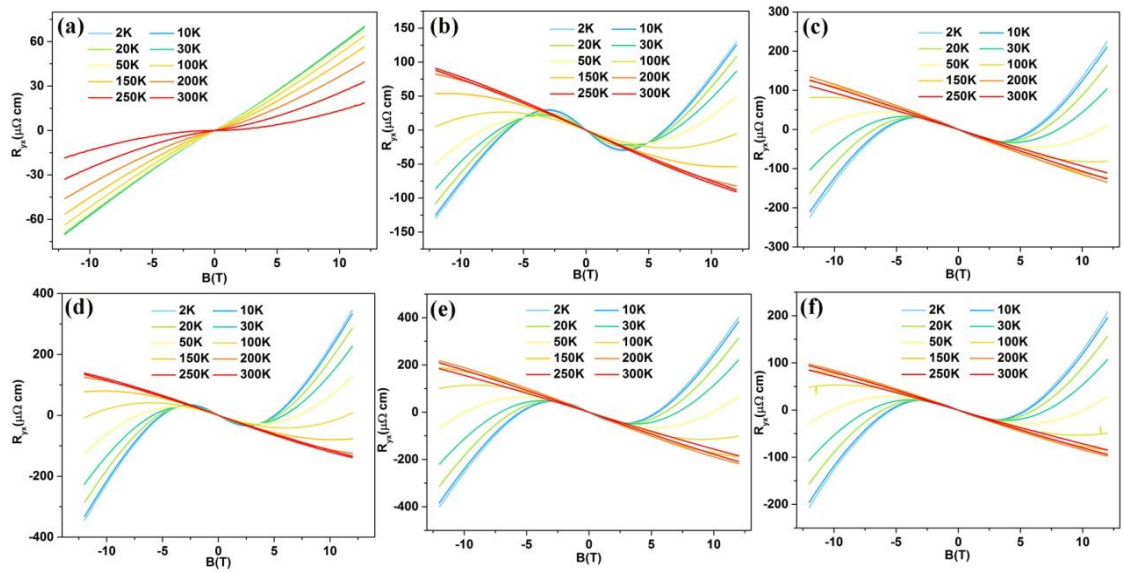

Figure S3. The Hall resistivity of the Mg$_3$Bi$_2$ series materials measured at different

temperatures in a magnetic field range between −12 and 12 Tesla.

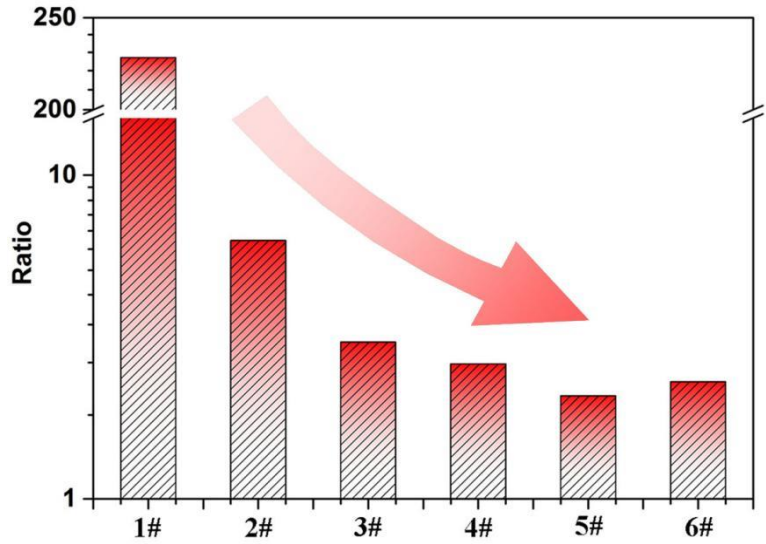

Figure S4. The ratio of carrier concentrations in $Mg_3Bi_2$ series samples.

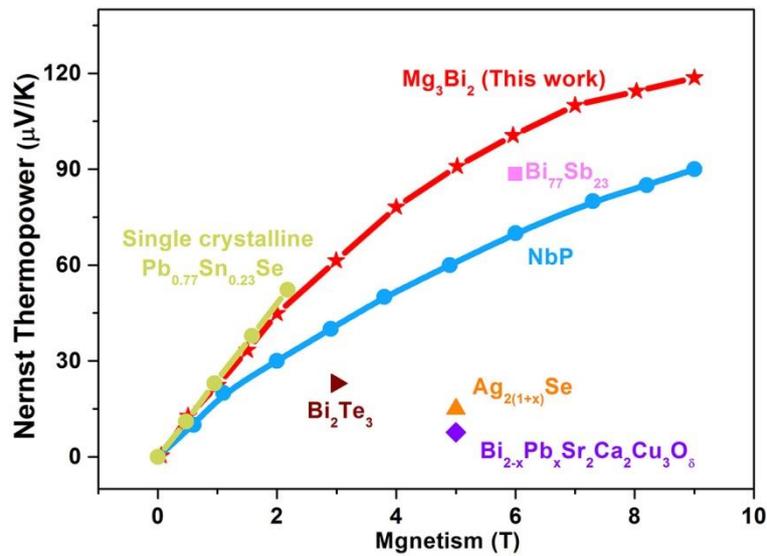

Figure S5. The comparison of Nernst (transverse) thermopower among polycrystalline $Mg_3Bi_2$, polycrystalline NbP [8], polycrystalline $Bi_{77}Sb_{23}$ [9], polycrystalline $Ag_{2(1+x)}Se$ [10], polycrystalline $Bi_{2-x}Pb_xSr_2Ca_2Cu_3O_\delta$ [11], single-crystal $Bi_2Te_3$ [12], and single-crystal $Pb_{0.77}Sn_{0.23}Se$ [13].

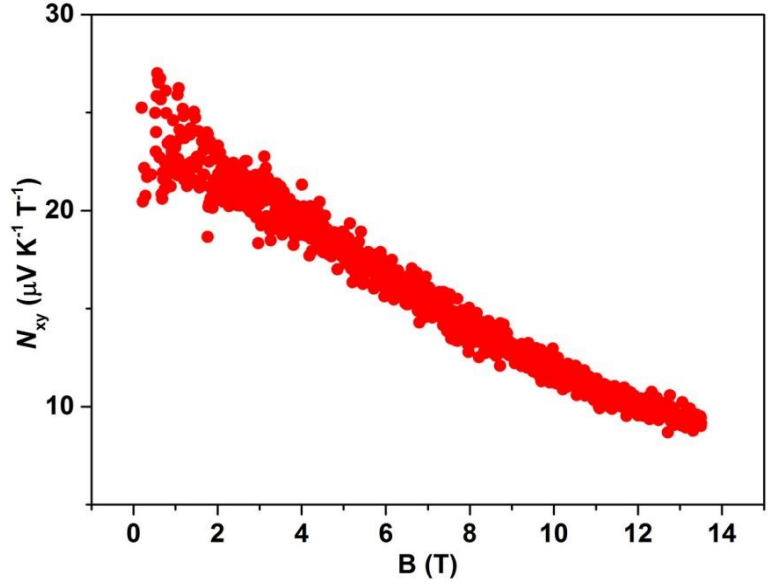

Figure S6. The magnetic field-dependent Nernst coefficient of the obtained $Mg_3Bi_2$ sample (Sample 5#) at 13.5 K.

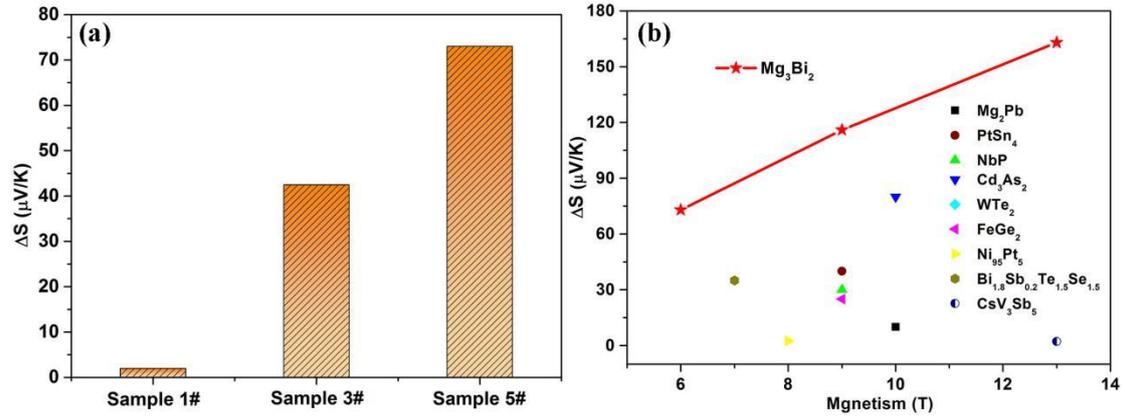

Figure S7. (a) The gain of longitudinal thermopower, defined as $\Delta S = S_{xx}(6\ \text{Tesla}) - S_{xx}(0\ \text{Tesla})$, in sample 1#, sample 3# and sample 5#; (b) The comparison of longitudinal thermopower gain $\Delta S$ between polycrystalline $Mg_3Bi_2$ and other MTE materials, such as polycrystalline NbP [8], single-crystal $Mg_2Pb$ [14], single-crystal $PtSn_4$ [15], single-crystal $Cd_3As_2$ [16], single-crystal $WTe_2$ [17], single-crystal $FeGe_2$ [18], single-crystal $Ni_{95}Pt_5$ [19], single-crystal $Bi_{1.8}Sb_{0.2}Te_{1.5}Se_{1.5}$ [20], and single-crystal $CsV_3Sb_5$ [21].

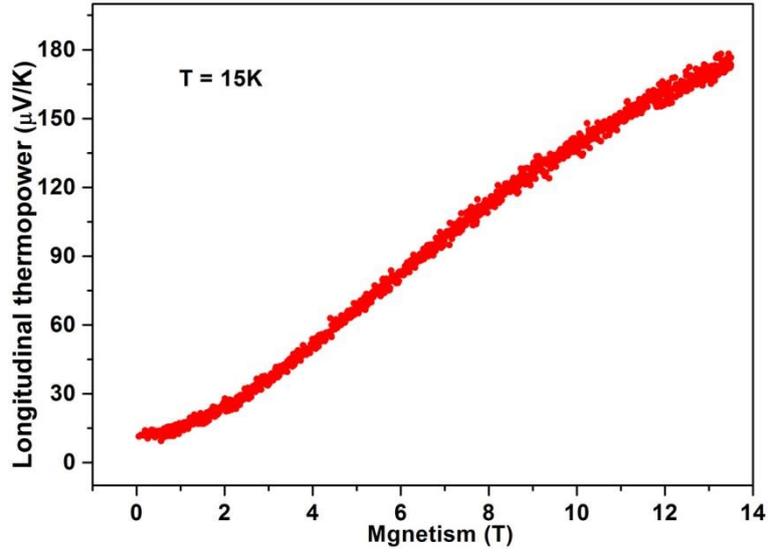

Figure S8. The magnetic field-dependent longitudinal thermopower for the obtained Mg$_3$Bi$_2$ (sample 5#) in the magnetic field range between 0 Tesla and 13 Tesla.

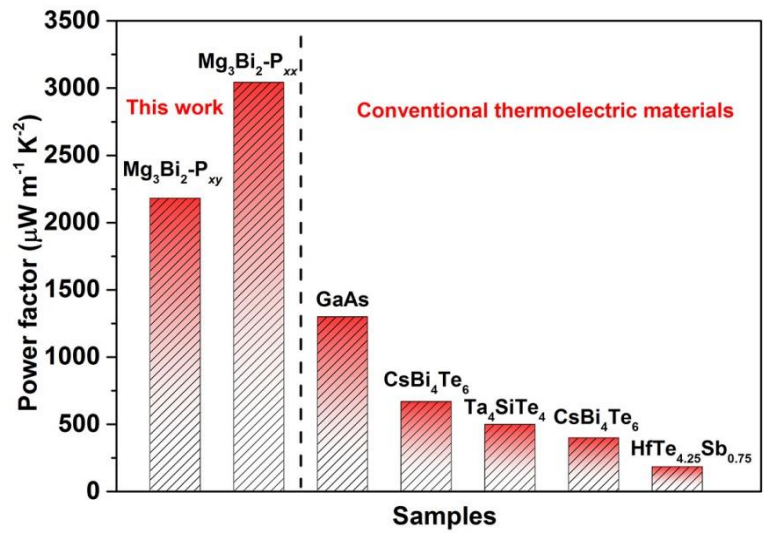

Figure S9. The comparison of power factor between MTE Mg$_3$Bi$_2$ sample and conventional thermoelectric materials at 15 K, such as GaAs [22], CsBi$_4$Te$_6$ [23, 24], Ta$_4$SiTe$_4$ [25], HfTe$_{4.25}$Sb$_{0.75}$ [26].

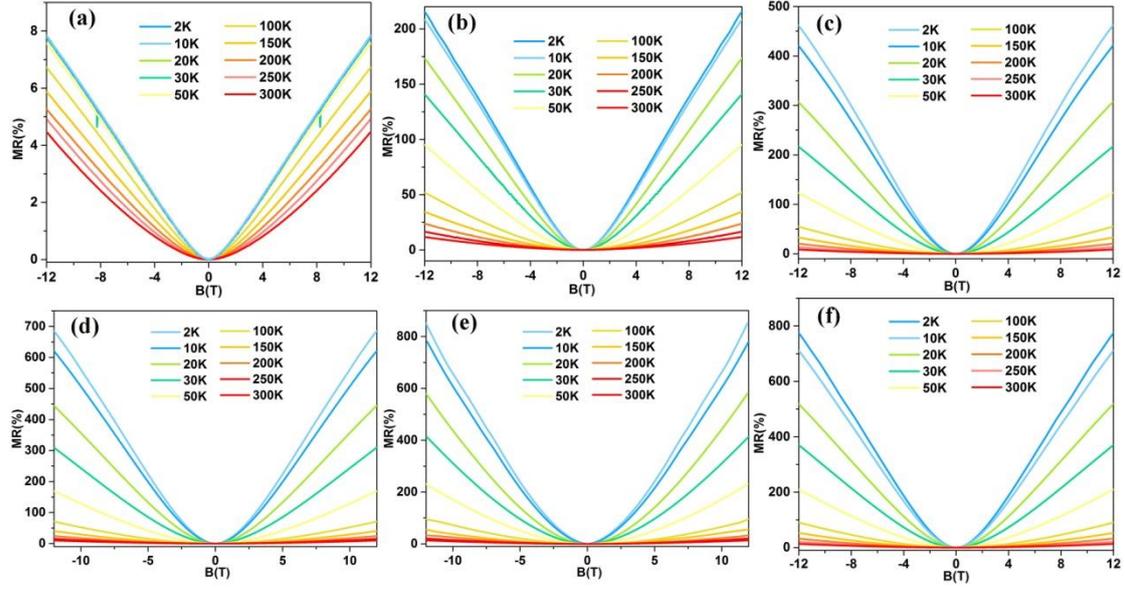

Figure S10. The magnetoresistance of $Mg_{3+x}Bi_2$ series samples with (a) $x = 0$, (b) $x = 0.05$, (c) $x = 0.1$, (d) $x = 0.2$, (e) $x = 0.3$, and (f) $x = 0.5$ at different temperatures in the magnetic field range between -12 Tesla and 12 Tesla.

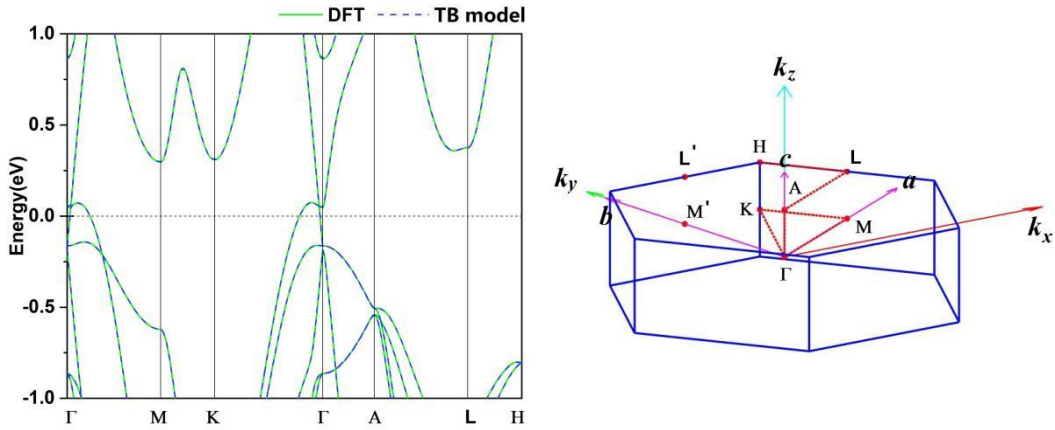

Figure S11. (left panel) DFT Calculated band structure (solid line) and the fitted band structure with tight binding model (dashed line) implemented in Wannier90 package without SOC for $Mg_3Bi_2$. (right panel) The first Brillouin zone, high symmetric $k$ points and the corresponding $k$-path (red dashed lines) for the band structure calculations.

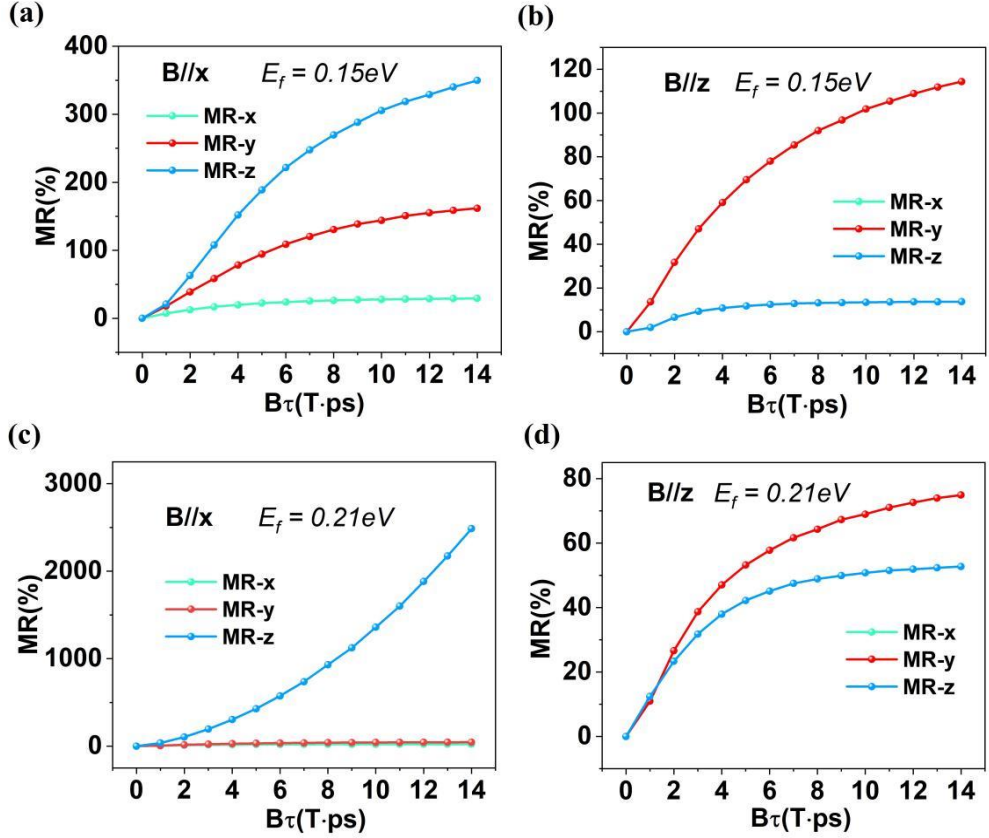

Figure S12. Field dependence of MR for $Mg_3Bi_2$ with SOC with Fermi level $E_f$ = 0.15 (a, b) and 0.21 (c, d) eV along the $x$ (cyan), $y$ (red), and $z$ (blue) directions with (a, c) **B**//$x$ and (b, d) **B**//$z$, respectively. We can see that for $E_f$ = 0.15 eV (a, b) the MR of $Mg_3Bi_2$ is saturated with **B** along both $x$ and $z$ directions, which is consistent with the closed feature of carrier orbits (Figure S11). When Fermi level is higher than 0.19 eV (e.g., $E_f$ = 0.21 eV), MR is suppressed compared with that of $E_f$ = 0.19 eV, due to the emergence of electron carriers with closed orbit around Γ point.

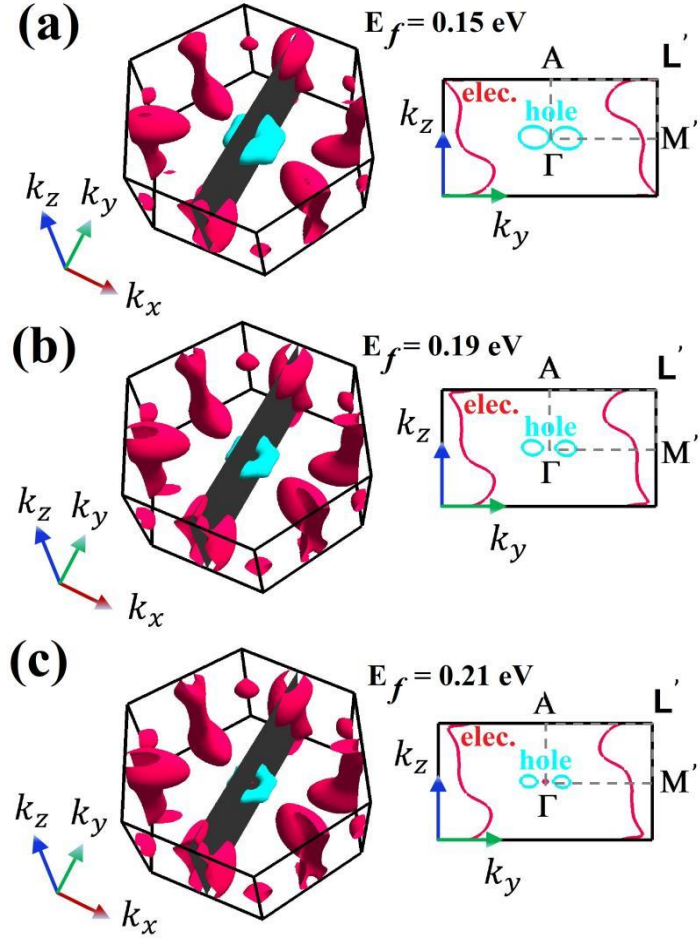

Figure S13. Fermi-surface contours with Fermi levels $E_f$ = 0.15 eV, 0.19 eV and 0.21 eV (a-c), respectively, and the corresponding cross sections of the Fermi surface produced by the $k_x = 0$ plane. We can see that for $E_f$ = 0.15 eV (a) both the electron and hole orbits are almost closed, indicating a saturated feature of MR. When Fermi level is higher than 0.19 eV, the electron carriers with closed orbit emergence around Γ point (c) leading to a suppression of MR.

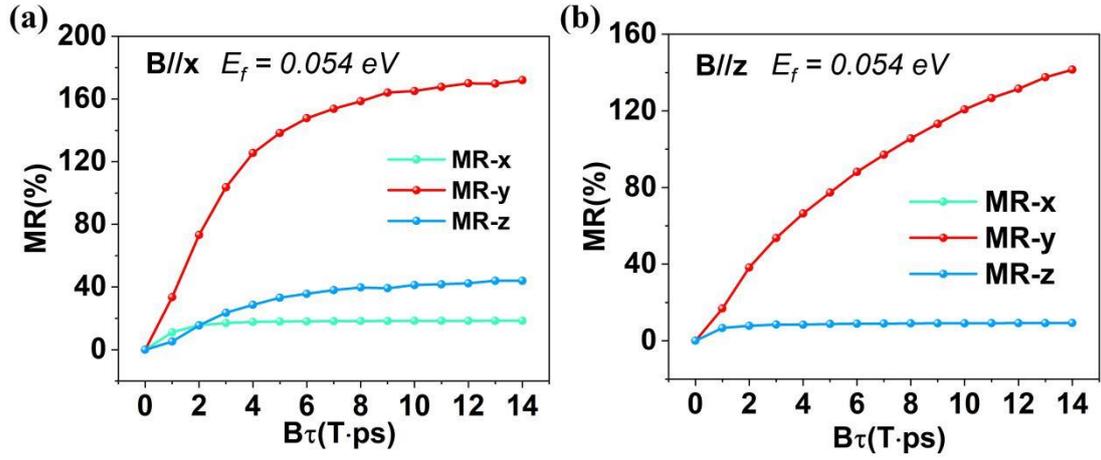

Figure S14. Field dependence of MR for $Mg_3Bi_2$ without SOC with Fermi level at the nodal line band crossing point (i.e., $E_f$ = 0.054 eV) along the *x* (cyan), *y* (red), and *z* (blue) directions with (a) $B//x$ and (b) $B//z$, respectively. We can see that MR for $Mg_3Bi_2$ are smaller and saturated for both cases compared with that of SOC.

Table S1. Nominal components of the Mg$_3$Bi$_2$ series of samples in the synthesis process.

| Sample | #1 | #2 | #3 | #4 | #5 | #6 |
|---|---|---|---|---|---|---|
| Nominal ratio | Mg$_3$Bi$_2$ | Mg$_{3.1}$Bi$_2$ | Mg$_{3.2}$Bi$_2$ | Mg$_{3.3}$Bi$_2$ | Mg$_{3.4}$Bi$_2$ | Mg$_{3.5}$Bi$_2$ |

Table S2. Effective mass (unit m$_e$) tensor of Mg$_3$Bi$_2$ at some high symmetric ***k*** points with SOC.

| ***k*** | xx | xy | xz | yx | yy | yz | zx | zy | zz |
|---|---|---|---|---|---|---|---|---|---|
| Γ(0.19) | 0.14 | 0 | 0 | 0 | 0.14 | 0 | 0 | 0 | 0.11 |
| Γ(0.14) | 0.08 | 0 | 0.03 | 0 | 0.08 | 0.03 | 0.03 | 0.03 | 4.73 |
| M | 0.48 | 0.20 | -0.58 | 0.20 | 0.24 | -0.34 | -0.58 | -0.34 | -1.44 |
| K | 0.38 | -0.01 | 0 | -0.01 | 0.37 | 0 | 0 | 0 | 0.21 |
| A-L | 0.39 | 0.13 | 0.20 | 0.13 | 0.24 | 0.11 | 0.20 | 0.11 | -0.16 |

We can see that at Γ point with $E_f$ = 0.19 eV the effective masses are all very small for three directions, indicating a large carrier mobility.